\documentstyle[12pt,epsfig,epsf]{article}
\catcode`\@=11

\@addtoreset{equation}{section}
\def\theequation{\arabic{equation}}

\def\@normalsize{\@setsize\normalsize{15pt}\xiipt\@xiipt
\abovedisplayskip 14pt plus3pt minus3pt%
\belowdisplayskip \abovedisplayskip
\abovedisplayshortskip  \z@ plus3pt%
\belowdisplayshortskip  7pt plus3.5pt minus0pt}
\def\small{\@setsize\small{13.6pt}\xipt\@xipt
\abovedisplayskip 13pt plus3pt minus3pt%
\belowdisplayskip \abovedisplayskip
\abovedisplayshortskip  \z@ plus3pt%
\belowdisplayshortskip  7pt plus3.5pt minus0pt
\def\@listi{\parsep 4.5pt plus 2pt minus 1pt
            \itemsep \parsep
            \topsep 9pt plus 3pt minus 3pt}}

\def\underline#1{\relax\ifmmode\@@underline#1\else
        $\@@underline{\hbox{#1}}$\relax\fi}
\@twosidetrue
\relax

\catcode`@=12

\catcode`\@=11

\def\section{\@startsection{section}{1}{\z@}{3.5ex plus 1ex minus
   .2ex}{2.3ex plus .2ex}{\large\bf}}

\def\ps@headings{\def\@oddfoot{}\def\@evenfoot{}
\def\@oddhead{\hbox{}\hfill
        \makebox[.5\textwidth]{\raggedright\ignorespaces --\thepage{}--
        \hfill }}
\def\@evenhead{\@oddhead}
\def\subsectionmark##1{\markboth{##1}{}}
}

\ps@headings

\catcode`\@=12

\relax

\def\figcap{\section*{Figure Captions\markboth
        {FIGURECAPTIONS}{FIGURECAPTIONS}}\list
        {Fig. \arabic{enumi}:\hfill}{\settowidth\labelwidth{Fig. 999:}
        \leftmargin\labelwidth
        \advance\leftmargin\labelsep\usecounter{enumi}}}
 \relax
\def\tablecap{\section*{Table Captions\markboth
        {TABLECAPTIONS}{TABLECAPTIONS}}\list
        {Table \arabic{enumi}:\hfill}{\settowidth\labelwidth{Table 999:}
        \leftmargin\labelwidth
        \advance\leftmargin\labelsep\usecounter{enumi}}}
 \relax
\def\reflist{\section*{References\markboth
        {REFLIST}{REFLIST}}\list
        {[\arabic{enumi}]\hfill}{\settowidth\labelwidth{[999]}
        \leftmargin\labelwidth
        \advance\leftmargin\labelsep\usecounter{enumi}}}
 \relax

\catcode`\@=11

\def\marginnote#1{}
\newcount\hour
\newcount\minute
\newtoks\amorpm
\hour=\time\divide\hour by60
\minute=\time{\multiply\hour by60 \global\advance\minute by-
\hour}
\edef\standardtime{{\ifnum\hour<12 \global\amorpm={am}%
    \else\global\amorpm={pm}\advance\hour by-12 \fi
    \ifnum\hour=0 \hour=12 \fi
    \number\hour:\ifnum\minute<100\fi\number\minute\the\amorpm}}
\edef\militarytime{\number\hour:\ifnum\minute<100\fi\number\minute}
\def\draftlabel#1{{\@bsphack\if@filesw {\let\thepage\relax
  \xdef\@gtempa{\write\@auxout{\string
    \newlabel{#1}{{\@currentlabel}{\thepage}}}}}\@gtempa
    \if@nobreak \ifvmode\nobreak\fi\fi\fi\@esphack}
     \gdef\@eqnlabel{#1}}
\def\@eqnlabel{}
\def\@vacuum{}
\def\draftmarginnote#1{\marginpar{\raggedright\scriptsize\tt#1}}
\def\draft{\oddsidemargin -.5truein
        \def\@oddfoot{\sl preliminary draft \hfil
        \rm\thepage\hfil\sl\today\quad\militarytime}
        \let\@evenfoot\@oddfoot \overfullrule 3pt
        \let\label=\draftlabel
        \let\marginnote=\draftmarginnote
   
\def\@eqnnum{(\theequation)\rlap{\kern\marginparsep\tt\@eqnlabel}%
\global\let\@eqnlabel\@vacuum}  }
\def\preprint{\twocolumn\sloppy\flushbottom\parindent 1em
        \leftmargini 2em\leftmarginv .5em\leftmarginvi .5em
        \oddsidemargin -.5in    \evensidemargin -.5in
        \columnsep 15mm \footheight 0pt
        \textwidth 250mmin      \topmargin  -.4in
        \headheight 12pt \topskip .4in
        \textheight 175mm
        \footskip 0pt
        
\def\@oddhead{\thepage\hfil\addtocounter{page}{1}\thepage}
        \let\@evenhead\@oddhead \def\@oddfoot{} \def\@evenfoot{} 
}
\def\titlepage{\@restonecolfalse\if@twocolumn\@restonecoltrue\onecolumn
     \else \newpage \fi \thispagestyle{empty}\c@page\z@
        \def\thefootnote{\fnsymbol{footnote}} }
\def\endtitlepage{\if@restonecol\twocolumn \else  \fi
        \def\thefootnote{\arabic{footnote}}
        \setcounter{footnote}{0}}  
\catcode`@=12
\relax

\def\ps@headings{\def\@oddfoot{}\def\@evenfoot{}
\def\@oddhead{\hbox{}\hfill
        \makebox[.5\textwidth]{\raggedright\ignorespaces --\thepage{}--
        \hfill }}
\def\@evenhead{\@oddhead}
\def\subsectionmark##1{\markboth{##1}{}}
}

\ps@headings

\relax

\def\firstpage#1#2#3#4#5#6{
\begin{document}
\def\beq{\begin{equation}} 
\def\eeq{\end{equation}} 
\def\bea{\begin{eqnarray}} 
\def\eea{\end{eqnarray}} 
\def\bq{\begin{quote}} 
\def\eq{\end{quote}}
\def\ra{\rightarrow} 
\def\lra{\leftrightarrow} 
\def\ups{\upsilon}
\def\bq{\begin{quote}} 
\def\eq{\end{quote}}
\def\ra{\rightarrow} 
\def\un{\underline}
\def\ov{\overline}
\newcommand{\cm}{Commun.\ Math.\ Phys.~}
\newcommand{\prl}{Phys.\ Rev.\ Lett.~}
\newcommand{\pr}{Phys.\ Rev.\ D~}
\newcommand{\pl}{Phys.\ Lett.\ B~}
\newcommand{\ibar}{\bar{\imath}}
\newcommand{\jbar}{\bar{\jmath}}
\newcommand{\np}{Nucl.\ Phys.\ B~}
\newcommand{\F}{{\cal F}}
\renewcommand{\L}{{\cal L}}
\newcommand{\A}{{\cal A}}
\def\154{\frac{15}{4}}
\def\153{\frac{15}{3}}
\def\32{\frac{3}{2}}
\def\254{\frac{25}{4}}
\begin{titlepage}
\nopagebreak
\title{\begin{flushright}
        \vspace*{-1.8in}
        {\normalsize CERN-TH/96-367}\\[-9mm]
        {\normalsize CPTH-S490.0197}\\[-9mm]
        {\normalsize IOA-TH/97-001}\\[-9mm]
        {\normalsize hep-ph/9701292}\\[4mm]
\end{flushright}
\vfill
{#3}}
\author{\large #4 \\[1.0cm] #5}
\maketitle
\vskip -7mm     
\nopagebreak 
\begin{abstract}
{\noindent #6}
\end{abstract}
\vfill
\begin{flushleft}
\rule{16.1cm}{0.2mm}\\[-3mm]
$^{\star}${\small Research supported in part by the EEC under the 
\vspace{-4mm} TMR contract ERBFMRX-CT96-0090 and by ${\Pi}$ENE${\Delta}$
91E${\Delta}$300.}\\ 
$^{\dagger}${\small Laboratoire Propre du CNRS UPR A.0014.}\\
January 1997
\end{flushleft}
\thispagestyle{empty}
\end{titlepage}}

\def\simlt{\stackrel{<}{{}_\sim}}
\def\simgt{\stackrel{>}{{}_\sim}}
\date{}
\firstpage{3118}{IC/95/34}
{\large\bf Unification Bounds on the Possible $N=2$ Supersymmetry-Breaking
Scale$^{\star}$} 
{I. Antoniadis$^{\,a,b}$, J. Ellis$^{\,b}$ and G.K. Leontaris$^{\,c}$}
{\normalsize\sl
$^a$Centre de Physique Th\'eorique, Ecole Polytechnique$^\dagger$,
{}F-91128 Palaiseau, France\\[-3mm]
\normalsize\sl
$^b$ CERN, TH Division, 1211 Geneva 23, Switzerland\\[-3mm]
\normalsize\sl
$^c$Theoretical Physics Division, Ioannina University,
GR-45110 Ioannina, Greece.}
{In this letter, the possible appearance of $N=2$ supersymmetry 
at a low energy scale is investigated in the context of unified theories.
Introducing mirror particles for all the gauge and matter multiplets
of the Minimal Supersymmetric extension of the Standard Model (MSSM),
the measured values of sin$^2 \theta_W$ and $\alpha_3(M_Z)$ 
indicate that the $N=2$ threshold
scale $M_{S_2}$ cannot be lower than $\sim 10^{14}$GeV. If the
$U(1)$ normalization coefficient $k$ is treated as a free parameter, $M_{S_2}$
can be as low as $10^9$ GeV. On the other hand, if mirror quarks and 
leptons are absent and a non-standard  value for $k$ is used, 
$N=2$ supersymmetry breaking could in principle occur at the electroweak
scale. }

It has recently been realized \cite{.} that $N = 2$ supersymmetry can be 
broken spontaneously to $N=1$ in the context of local quantum field theory,
which opens up the possibility that $N=2$ supersymmetry may
become relevant at some intermediate energy scale below the Planck or
string scale.
Possible $N=2$
extensions of the Standard Model (SM) have been studied in the past \cite{f}  
and they are much more restrictive than the $N=1$ framework. In
particular, because of the vanishing of $Str(M^2)$ after supersymmetry
breaking, they guarantee the absence of 
all field-dependent
quadratic divergences in the scalar potential which, is a desirable
ingredient for
solving the hierarchy problem. 
In this letter, we derive lower bounds on the $N=2$
breaking scale in the context of unified theories.

It is well known that the $N=1$ supersymmetric beta-function coefficients $b_i$
allow the three gauge couplings of the electroweak and strong forces to
attain a common value at a scale $M_X\sim 10^{16}$ GeV. If $N=2$
supersymmetry appears at some intermediate threshold scale $M_{S_2}$, 
the beta-function coefficients
change drastically due to the contributions of the $N=2$ superpartners of
all the SM states. In terms of $N=1$ superfields, these are one adjoint for
each group factor of the gauge symmetry, and one mirror (of opposite
chirality) for each matter field. The introduction of mirrors for both
Higgs doublets is also necessary for the breaking of the $SU(2)$ gauge
symmetry. As a result, gauge coupling
unification occurs in general at a different scale $M_U$, which turns out
to be greater than $M_X$.

In this letter, we study the allowed values of $M_U$ and the corresponding
lower bounds for the $N=2$ scale $M_{S_2}$, which are consistent with 
the low-energy data. We find an interesting correlation between the
two scales, namely that higher $M_U$ implies lower $M_{S_2}$.
Fixing the normalization of the
$U(1)$ hypercharge to the standard value $k=5/3$ we find that $M_{S_2}$
cannot be smaller than $\sim 10^{14}$ GeV. However, if a
different $U(1)$ hypercharge normalization is allowed, $M_{S_2}$ can be as
low as $\sim 10^{9}$ GeV.

In the energy range between $M_{S_2}$ and the unification scale $M_U$,
the beta-function coefficients read:
\bea
b_1^{N=2}= \frac{66}5\ , & b_2^{N=2}= 10\ , & b_3^{N=2}= 6\ ,
\label{n2}
\eea
for the $U(1)$, $SU(2)$ and $SU(3)$ gauge group factors respectively.
For simplicity, we assume that $N=1$ supersymmetry remains exact down to the
$M_Z$ scale, so that in the range $M_Z$ to $M_{S_2}$ the
beta-function
coefficients are those of the $N=1$ Minimal Supersymmetric 
extension of the Standard Model (MSSM), namely:
\bea
b_1^{N=1}= \frac{33}5\ , & b_2^{N=1}= 1\ , & b_3^{N=1}= -3\ .
\eea
Using the renormalization-group equations for the three gauge couplings, 
we first eliminate the $M_{S_2}$ scale to obtain the following formula 
for the unification scale $M_U$ in terms of the experimentally-measured 
low-energy parameters:
\bea
\log{\frac{M_U}{M_Z}}={\pi\over 2\alpha}\left(
\sin^2\theta_W - \frac{\alpha}{\alpha_3}\right) \ ,
\label{sin2}
\eea
where $\alpha$, $\alpha_3$ are the low energy electromagnetic and strong
coupling constants, respectively. 

In this paper, we make the self-consistent approximation
of ignoring
low-energy thresholds, two-loop effects in the region below $M_{S_2}$,
and the model-dependent high-energy threshold around $M_{S_2}$.
Above this scale, $N=2$ supersymmetry is unbroken and there are no higher 
loop corrections. We should point out that the effects 
we ignore are potentially important and may
alter our results. It is known that these are important
for detailed comparisons of $N=1$ supersymmetric GUTs with the available
experimental
data (for a review, see~\cite{je}). However, since the high-energy
threshold effects are currently unknown, we
prefer to restrict this analysis to the self-consistent
one-loop approximation, and add larger theoretical error bars to the
purely experimental errors on the low-energy value of
$\sin^2\theta_W$.

The experimental values of the low-energy parameters that we use as the
basis for our
determination of $M_U$ are~\cite{b}:
\bea
\sin^2\theta_W=0.2316\pm 0.0004\, (\pm 0.003)\qquad
\alpha_3=0.118\pm 0.005\ ,
\label{values}
\eea
where the second error in $\sin^2\theta_W$ accounts for the theoretical
uncertainties mentioned above, and has been chosen to have the same
magnitude as the two-loop effect in the desert in conventional $N=1$
unification.
The resulting $M_U$ region is shown in Fig.~1. The dashed lines
represent the first (experimental) error in $\sin^2\theta_W$ of 
eq.~(\ref{values}). We have also indicated the effect of relaxing the
experimental constraints on $\alpha_3$, allowing it to vary over the
range $\sim 0.11-0.13$. We deduce that,
despite the introduction of the new free parameter representing the $N=2$
threshold scale, the
low-energy data give a rather stringent constraint on the unification
mass, which has to be  less than 
$2 \times 10^{17}$ GeV. On the other hand, the assumed hierarchy
of scales, $M_{S_2}\le M_U$, implies the constraint:
\bea
\sin^2\theta_W\ge {3\over 8}-{7\alpha\over 4\pi}\ln{M_U\over M_Z}\ ,
\label{constr}
\eea
which requires $M_U\simgt 10^{16}$ GeV. The constraint (\ref{constr}) is
represented in Fig.~1 by a straight line, corresponding to $M_{S_2}= M_U$, 
which excludes values of
($\sin^2\theta_W , M_U$) below it and to its left.

\begin{figure}
\begin{center}
\leavevmode
\epsfig{figure=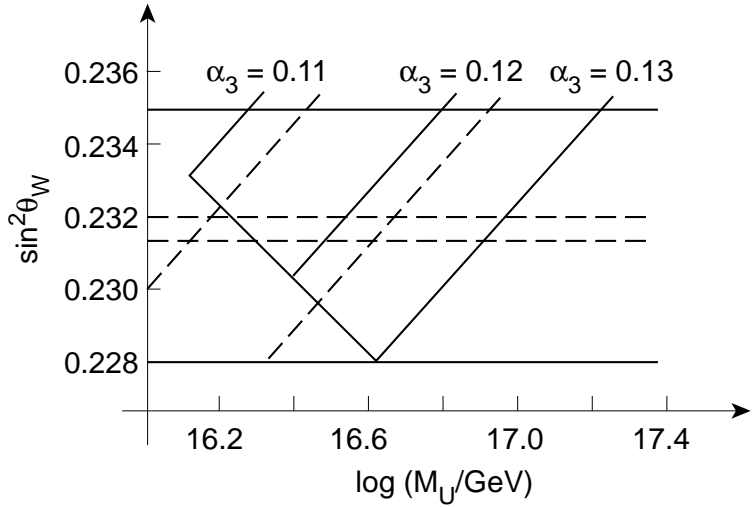,width=10cm}
\caption[]{ The range of $M_U$ allowed by the experimental values 
of $\alpha_3$ and $\sin^2\theta_W$
given in eq.~(\ref{values}). Dotted lines correspond to 
purely experimental errors, whilst the solid ones include 
an allowance for theoretical
uncertainties. }
 \end{center}
\end{figure}

\begin{figure}
\begin{center}
\leavevmode
\epsfig{figure=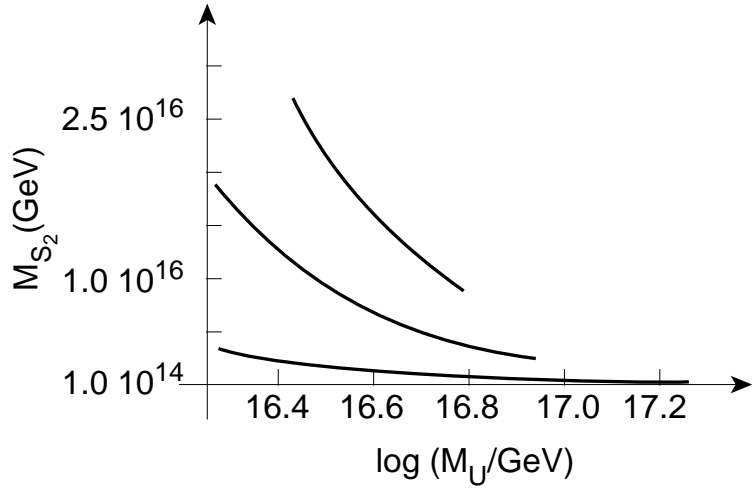,width=10cm}
\caption[]{ 
The $N=2$ scale $M_{S_2}$, as a function of
$M_U$, for three values of the minimal $N=1$ supersymmetric unification
scale: $M_X= (0.9, 1.86, 2.69)\times 10^{16}$ GeV.}
\end{center}
\end{figure}

We now come to the computation of
the intermediate $N=2$ scale. It is useful to express it as a function
of the parameters $M_U$ and $\sin^2\theta_W$, so that we can determine
its range in the parameter space of Fig.~1. We obtain
\bea
M_{S_2}=e^{\frac{4\pi}{3\alpha}\left(\frac 38-\sin^2\theta_W\right)}
          \left(\frac{M_Z}{M_U}\right)^{\frac 43}M_Z\ .
\label{MS2}
\eea
This expression should be compared with the one obtained for minimal 
supersymmetric 
grand unification scenario assuming that the only light particles are
those of the MSSM, where the unification
scale $M_X$ is given in the one-loop approximation by
\bea
M_X =e^{\frac{4\pi}{7\alpha}\left(\frac 38-\sin^2\theta_W\right)} M_Z\ .
\label{MX}
\eea
Using eq.~(\ref{MX}), we can rewrite $M_{S_2}$ in eq.~(\ref{MS2})
as:
\beq
M_{S_2} =  \left(\frac{M_X}{M_U}\right)^{\frac 73}M_U\ .
\label{MS2X}
\eeq
Thus, for a given $M_X$, or equivalently $\sin^2\theta_W$,
we can plot $M_{S_2}$ as a function of the unification scale
$M_U$ whose range was shown in Fig.~1. This is shown in Fig.~2.
We conclude that, when $M_U$ is near its lower bound $\sim 10^{16}$ GeV
$\sim
M_X$, then $M_{S_2}\sim M_X$, while as $M_U$ approaches its higher bound 
$\sim 2\times 10 ^{17}$ GeV then $M_{S_2}\sim 10^{14}GeV$.
 
It is important to note that, over the entire allowed $M_U$ range, the
value of the
gauge coupling at the unification scale  remains small: $\alpha_U \ll 1$.
In Table 1, we display the value of $\alpha_U$ for three representative
cases.
\begin{table}
\centering
\begin{tabular}{|c|c|c|c|} \hline
$\alpha_3$ & $M_U$/GeV &$M_{S_2}$/GeV& $1/\alpha_U$ \\ \hline
0.11  & $1.91\cdot 10^{16}$& $2.90\cdot 10^{15}$ &22.14\\
0.12  & $6.30\cdot 10^{16}$& $5.76\cdot 10^{14}$ &$17.92$\\
0.13  & $1.70\cdot 10^{17}$& $1.56\cdot 10^{14}$ &$14.46$
\\ \hline
\end{tabular}
\caption{ Lower bounds on $M_{S_2}$ and the corresponding values of
$M_U$ and $\alpha_U$, for three indicative choices of $\alpha_3$.}
\label{table:I}
\end{table}

As we show below, the scale $M_{S_2}$ could decrease if the $U(1)$ 
normalization coefficient $k$ is larger than its standard value $5/3$ 
at the unification mass. Conventional $N=1$
string unification needs small $k$ values to reconcile the high string
scale
with the low-energy value of the weak mixing angle $\sin^2\theta_W$. 
On the other hand, such non-standard $U(1)$ normalizations have been
discussed in the context of superstring models \cite{x}, which offer
the possibility that the $k$ parameter might be
larger than $5/3$. This is possible, for example,
if the hypercharge generator corresponds
to a linear combination of $U(1)$ factors,
with an embedding into a
higher-rank non-abelian gauge group. Such higher-level
constructions have been motivated by two phenomenological
considerations: they could guarantee the absence of 
color-singlet states with
fractional electric charges in four dimensional string models 
\cite{s} (though these could also be confined by
hidden-sector interactions, analogously to quarks in QCD~\cite{conf}).

A key observation is that eq.~(\ref{sin2}), which gives the unification
scale $M_U$, is independent of the normalization coefficient $k$. 
On the other hand, eq.~(\ref{MS2}), which gives the $M_{S_2}$ scale,
becomes for arbitrary values of $k$:
\bea
M_{S_2}   &=& \left[e^{\frac{4\pi}{3 \alpha}
\left(\frac{3}{8}-\frac{5}{6}\sin^2\theta_W\right)}
\left(\frac{M_Z}{M_U}\right)^{\frac {22}9}\right]^{-\frac{k-5/3}{k-11/9}}
\times
M_{S_2}{\vert_{}}_{k={5\over 3}}\ ,
\label{MS2k}
\eea
where  we have replaced the expression for $M_{S_2}$
in the $k=5/3$ case from eq.~(\ref{MS2}). It is clear that, as $k$
increases to values larger than $5/3$, the scale $M_{S_2}$ decreases
rapidly due to the exponential suppression factor in eq.~(\ref{MS2k}). 

In addition to the above expression for $M_{S_2}$, the formula for the 
gauge coupling at $M_U$, $\alpha_U$, also depends on $k$:
\bea
k &=& \frac{11}9+
 \frac{\alpha_U}{\alpha}\left(
  1+\frac{22}9\frac{\alpha}{\alpha_3}-\frac{14}3\sin^2\theta_W\right)
\label{kN2}
\eea
Requiring $\alpha_U\le 1$, we can thus obtain an upper bound on $k$ for any
given unification mass. In Table 2, we present these upper bounds for three
indicative values of the strong coupling $\alpha_3$. We see that, even
allowing for a larger value of $k$,
$M_{S_2}$ cannot be lower than about $\sim (10^8-10^9)$ GeV.
\begin{table}
\centering
\begin{tabular}{|l|c|r|c|c|} \hline
$\alpha_3$ & $M_U$/GeV & $k$&$M_{S_2}$/GeV& $\alpha_U$ \\ \hline
0.11 & $1.35\cdot 10^{16}$& 12.17 & $9.25\cdot 10^8$    & 1 \\
       & $2.00\cdot 10^{16}$& 10.97 & $1.21\cdot 10^9$    &   \\
0.12 & $2.70\cdot 10^{16}$& 11.95 & $2.60\cdot 10^9$    & 1 \\
       & $5.91\cdot 10^{16}$&  9.30 & $3.98\cdot 10^9$    &   \\
0.13 & $4.26\cdot 10^{16}$& 11.80 & $5.18\cdot 10^{9} $ & 1 \\
       & $1.71\cdot 10^{17}$&  7.65 & $1.30\cdot 10^{10}$ &   \\
\hline
\end{tabular}
\caption{Bounds on $k$ and $M_{S_2}$ for three choices of
$\alpha_3$, enforcing $\alpha_U=1$.}
\label{table:II}
\end{table}

At this point, one may ask whether $k$ can be large enough to
be able to impose charge quantization
without invoking confinement in the
hidden sector \cite{conf}. For this, one needs $k\ge 17/3$ \cite{s}.
It is obvious from table 2, that the answer to this question is 
positive provided that the unification coupling  is $\sim {\cal O}(1)$.
In Table 3 we give the $N=2$ scale  and $\alpha_U$ for the
next two allowed $k$ values consistent with  the charge
quantization condition. For $k=17/3$, the highest $\alpha_U$ value 
obtained is $\sim 0.68$.  We further observe that for
 $k=29/3$ the unification
coupling can reach the value $a_U=1$, which corresponds to the self-dual
point of the S-duality transformation: $a_U\to 1/a_U$, for a relatively
wide range of the
unification mass: $M_U\sim 5\times 10^{16} - 10^{17}$ GeV.

\begin{table}
\centering
\begin{tabular}{|c|c|c|c|c|l|} \hline
$\alpha_3$ &$\sin^2\theta_W$& $M_U$/GeV & $k$&$M_{S_2}$/GeV& $\alpha_U$ \\
 \hline
0.118 &0.2316& $2.52\cdot 10^{16}$& 17/3 & $5.32\cdot 10^9$    & 0.428 \\
0.130 &0.2350& $4.26\cdot 10^{16}$& 17/3 & $1.35\cdot 10^{10}$ & 0.420\\
0.130 &0.2280& $1.63\cdot 10^{17}$& 17/3 & $1.17\cdot 10^{10}$ & 0.679\\
\hline
0.118 &0.2316& $2.52\cdot 10^{16}$& 29/3 & $2.46\cdot 10^9$    & 0.814 \\
0.130 &0.2316& $8.67\cdot 10^{16}$& 29/3 & $8.25\cdot 10^9$    & 1.0 \\
0.118 &0.2348& $4.80\cdot 10^{16}$& 29/3 & $3.22\cdot 10^9$    & 1.0 \\
\hline
\end{tabular}
\caption{The $M_{S_2}$ scale and the value of the unification
coupling  $\alpha_U$
for two choices of
the normalization constant $k$. }
\label{table:IIa}
\end{table}

The reason that $\alpha_U$ becomes strong before the
$M_{S_2}$ scale can be lowered considerably
is essentially the large positive contribution to
the beta functions from all the extra $N=2$ superpartners, which include in
particular the
mirrors of the conventional quarks and leptons. The existence of the
latter is of course
problematic, since it is difficult to invent a mechanism which gives them
masses and at the same time generates chirality together with partial
supersymmetry breaking. Some examples overcoming this difficulty have been 
discussed in the context of string theory and/or using compactifications 
involving constant magnetic fields \cite{str}. These examples
suggest that it might be possible for
the mirror fermions to
form massive pairs with the Kaluza-Klein excitations,
whose spectrum is shifted by
the symmetry breaking. In
these cases, the
$N=2$ scale is linked to the compactification radius of an extra
dimension,
and one needs special
models with no large thresholds in order to be able to continue the
renormalization-group equations above $M_{S_2}$ \cite{nlt}.

In order to cover this possibility, we now repeat our analysis assuming no
mirrors for the known
chiral fermions (quarks and leptons). The beta-function coefficients then
read:
\bea
\tilde{b}_1^{N=2}= \frac{36}5, &\tilde{b}_2^{N=2}= 4, 
&\tilde{b}_3^{N=2}= 0\, .
\label{n2p}
\eea
We note that the differences $(\tilde{b}^{N=2}_i-\tilde{b}^{N=2}_j)$
remain
the same as the $(b^{N=2}_i-b^{N=2}_j)$ of eq.~(\ref{n2}).
Consequently, the $k$-independent expression (\ref{sin2})  for the
$M_U$  scale still holds. However, the relation (\ref{MS2k}) %
for $M_{S_2}$ is modified to become:
\bea
\tilde{M}_{S_2}   &=& \left[e^{\frac{4\pi}{3 \alpha}
\left(\frac{3}{8}-\frac{1}{2}\sin^2\theta_W\right)}
\left(\frac{M_Z}{M_U}\right)^{\frac 83}\right]^{-\frac{k-5/3}{k-1/3}}
\times
M_{S_2}{\vert_{}}_{k={5\over 3}}\ ,
\label{tMS2k}
\eea
where  again we used the expression for $M_{S_2}$
in the $k=5/3$ case from eq.~(\ref{MS2}).
In addition, the relation (\ref{kN2}) is modified as follows
\bea
k &=& \frac{1}3+
 \frac{\alpha_U}{\alpha}\left(
  1+\frac{8}3\frac{\alpha}{\alpha_3}-4\sin^2\theta_W\right)
\label{kN2p}
\eea
It is easy to see now that $k$ is allowed in principle
to attain values much larger than previously, whilst keeping $\alpha_U$ in 
the perturbative region. Moreover, the scale $\tilde{M}_{S_{2}}$  can      
be arbitrarily small even for  moderate values of $k$. 
 Now, eq.~(\ref{tMS2k}) 
provides an upper bound for $k$, based on the
phenomenological requirement
that $M_{S_{2}}$ cannot be lower than the weak scale $M_Z$:
\beq
k\le 3+ \frac{\alpha_3}{\alpha}\left(1-4\sin^2\theta_W\right)\ .
\label{kmax}
\eeq
Thus, from (\ref{sin2}), we find the upper bound  $k\simlt 4.24$, 
attained when $\alpha_U\simlt 0.11$.
Note that on the
boundary $\tilde{M}_{S_{2}}=M_Z$ one has $\alpha_U=\alpha_3$, since the
beta function (\ref{n2p}) of $SU(3)$ vanishes. Moreover, uncertainties
from two-loop corrections in this latter case are eliminated, as the
$N=2$ scale remains down to $M_Z$. 
Note also that, unlike the previous case, the present bound on $k$
is smaller that the minimum value $k=17/3$ required from
the charge quantization condition.

If in addition to omitting mirrors of the quarks and leptons, we
also assume there are no
mirrors for the Higgses, we are left to consider only the effect of
adjoint
matter ($SU(3)$ octets and $SU(2)$ triplets) at some intermediate scale.
This possibility has been considered previously with the aim of increasing
the unification mass close to the string scale \cite{bfy}.

In conclusion: in the context of unified models having as effective
low-energy theory the minimal supersymmetric extension of the Standard
Model, we have 
derived bounds on a possible $N=2$ supersymmetry-breaking scale.
Assuming the presence of mirror partners for all the chiral matter and
Higgs fields and assuming the canonical normalization of the $U(1)$
of hypercharge, we
have 
found that the $N=2$ scale cannot be lower than $10^{14}$ GeV. On the
other hand, if one allows
a non-standard $U(1)$ normalization, the $N=2$ scale could be as low as
$10^9$ GeV. If there are no mirrors for quarks and
leptons, the $N=2$ breaking scale could be as low as the electroweak
scale, but there are still significant restrictions on the normalization
of the $U(1)$ of hypercharge.

\vspace*{1cm}

{\bf Acknowledgement.} I.A. would like to thank the Theoretical 
Physics Division  of Ioannina University for kind hospitality.
 G.K.L. would like to thank the Centre de
Physique Th\'eorique
of the \'Ecole Polytechnique and Theory Division of CERN 
for  kind hospitality while part of
this work was being done.



\begin{thebibliography}{99}

\bibitem{.} I. Antoniadis, H. Partouche and T.R. Taylor, \pl 372 (1996) 155;\\
S. Ferrara, L. Girardello and M. Porrati, \pl 376 (1996) 275;\\
J. Bagger and A. Galperin, hep-th/9608177.
\bibitem{f} P. Fayet, \pl 142 (1984) 263; \pl 159 (1985) 121.
\bibitem{je} J. Ellis, Rapporteur Talk at the {\it
International Symposium on Lepton and Photon Interactions
at High Energies}, Beijing 1995, hep-ph/9512335.
\bibitem{b} W. de Boer, Review presented at the {\it 28th 
International Conference on High-Energy Physics}, Warsaw
1996, hep-ph/9611395;\\ P.N. Burrows, 
Review presented at the {\it International Symposium on Radiative
Corrections}, Krakow 1996, hep-ex/9612007.
\bibitem{x} L. Ib\'a\~nez, \pl 318 (1993) 73;\\
K.R. Dienes, A. Faraggi and J. March-Russell, \np 467(1996)44.
\bibitem{s} A. N. Schellekens, \pl 237 (1990) 363.
\bibitem{conf} I. Antoniadis, J. Ellis, J. Hagelin and D.V. Nanopoulos,
\pl 231 (1989) 65;\\
J. Ellis, J.L. Lopez and D.V. Nanopoulos, \pl 245 (1990) 375.
\bibitem{str} I. Antoniadis, K. Benakli and M. Quir\'os, \pl 331 (1994) 313;\\
C. Bachas, hep-th/9503030.
\bibitem{nlt} I. Antoniadis, \pl 246 (1990) 377;\\
E. Kiritsis, C. Kounnas, P.M. Petropoulos and J. Rizos, \pl 385 (1996) 87.
\bibitem{bfy} C. Bachas, C. Fabre and T. Yanagida, \pl 370 (1996) 49;\\
M. Bastero-Gil and B. Brahmachari, hep-ph/9610374.
\end{thebibliography}
\end{document}